\documentclass[prd,superscriptaddress,nofootinbib,11pt, tightenlines]{revtex4-2} 
\usepackage{epsfig}  
\usepackage{bbold}
\usepackage{amsmath} 
\usepackage{amsfonts}  
\usepackage{xfrac}
\usepackage{float}    
\usepackage{amssymb}    
\usepackage{slashed} 
\usepackage{color}      
\usepackage{pbox}
\usepackage{bbding}
\usepackage{caption}
\usepackage{subcaption}
\usepackage{ragged2e}
\usepackage{array,multirow,makecell}
\usepackage[colorlinks,citecolor=blue, linkcolor = blue, urlcolor = blue]{hyperref}
\usepackage{cleveref}
\usepackage{titlesec} 
\usepackage{scrextend}
\usepackage{adjustbox}
\usepackage{natbib} 
\usepackage{lipsum}
\usepackage{relsize}
\usepackage[normalem]{ulem} 
\usepackage{xcolor,hyperref}
\usepackage{cancel}
\usepackage{marginnote}
\usepackage{contour}
\usepackage{ulem}
\usepackage{tikz}
\usepackage[compat=1.1.0]{tikz-feynman}
\usetikzlibrary{external}
\usepackage{orcidlink}

\newcommand{\be}{\begin{equation}} 
\newcommand{\ee}{\end{equation}} 
\newcommand{\bea}{\begin{eqnarray}}
\newcommand{\eea}{\end{eqnarray}}
\newcommand{\bee}{\begin{eqnarray*}}
	\newcommand{\eee}{\end{eqnarray*}}

\newcommand{\mchi}{m_{\chi}}
\newcommand{\nnu}{\nonumber\\}

\newcommand{\KeV}{{\, \rm KeV}}
\newcommand{\MeV}{{\, \rm MeV}}
\newcommand{\GeV}{{\, \rm GeV}}
\newcommand{\TeV}{{\, \rm TeV}}

\newcommand{\lamchiH}{\lambda_{\chi H}}
\newcommand{\lamchiD}{\lambda_{\chi\Delta}}
\newcommand{\lamHD}{\lambda_{H\Delta}}

\begin{document}

\date{\today}

\title{Analysis of Freeze-in scenario with a scalar Leptoquark and a scalar Dark Matter } 
%
\author{Joydeep Roy\orcidlink{0000-0001-7406-0127}}
\email{joydeeproy@acharya.ac.in}
\email{joyroy.phy@gmail.com}
\affiliation{School of Physical Sciences, Indian Association for the Cultivation of Science, Jadavpur, Kolkata 700032, India}
\affiliation{Department of Physics, Acharya Institute of Technology, Bengaluru, 560107, Karnataka, India}

\begin{abstract}

  Dark Matter relic density generation through \textit{freeze-in} mechanism where dark matter particles interact feebly with visible sector particles, is an alternative approach to well-studied and most popular \textit{freeze-out} paradigm. We study this \textit{freeze-in} scenario in the presence of a scalar leptoquark that interacts with both dark matter and Standard Model particles with renormalizable interactions. We discuss the effect of the presence of such a heavy particle, a scalar leptoquark with mass $\geq 1.5 \TeV$, in the thermal bath and on subsequent relic density generation. We explore the parameter space of such a framework, consisting of two masses and three dimensionless couplings. We numerically study the interaction rates and relic density as functions of these parameters and determine their values consistent with the dark matter constraints.
\end{abstract}

\maketitle
\newpage


\section{Introduction}

One of the major directions of particle physics research in recent times is strongly motivated by the goal of exploring the nature and properties of Dark Matter (DM), which is nonluminous, non-baryonic matter that accounts for almost $85\%$ of the matter density of the whole Universe \cite{Planck:2015fie}. Several cosmological and astrophysical evidence, such as the growth of structure and the cosmic evolution, dynamical observations of stars and galaxies, gravitational lensing measurements or big-bang nucleosynthesis have proven the existence of DM with certainty, but they can not provide a detailed nature of it. Among many other things, we still don’t know their mass, whether they are fundamental or composite particles or whether they are of scalar, vector or fermionic in nature etc.. These inadequacy of our knowledge have driven particle physicists to come up with the idea of Weakly Interacting Massive Particles (WIMPs) paradigm \cite{Steigman:1984ac}  that has dominated most of the theoretical and experimental works in DM searches till now. The outcome of this setup is the observed DM relic density known as the WIMP \textit{miracle} \cite{Arcadi:2017kky}.

Knowledge of standard cosmology, with the assumption of a radiation-dominated (RD) universe, tells us that, to produce the Cosmic Microwave Background (CMB) spectrum as we see today as well as the current number density of observed elements, the Standard Model (SM) particles must have been in thermal equilibrium with dark sector particles. Also, the energy density of those SM particles must be the dominant force behind the energy evolution of the Universe from the time of the Big Bang.  Considering these criteria, for a particle under the WIMP scenario to be in thermal equilibrium with the early universe and then gradually produce the observed relic density at present time, it can take any mass value from $\KeV$ to $\TeV$. Moreover, to generate the observed dark matter density, a self-annihilating DM must have a cross-section $\sigma v \sim 10^{-26} \rm ~ cm^3/s$ which is similar to the result that we get from electroweak force calculations. These features lead to extensive, vast and diverse experimental searches for a WIMP (or WIMP-like) DM particles, but to date no such experimental evidence, either from direct or indirect searches, has been observed. 

In contrast to the WIMP scenario, the DM doesn’t always have to be part of the same heat bath as that of the SM particles.  This idea of generation of DM abundance from the out-of-thermal-equilibrium setup is known as the \textit{freeze-in} scenario \cite{McDonald:2001vt,Kusenko:2006rh,Petraki:2007gq,Hall:2009bx}, which will be our chosen framework in this analysis. The basic features of this freeze-in scenario are the DM particles known as Feebly Interacting Massive Particles (FIMPs), which earn such a name due to their very weak couplings with the SM particles. Due to this feature, the DM particles are absent at the early stage of the Universe and can never be in thermal equilibrium with SM particles. Rather, they are generated by the scattering and decay of some particles that maintain thermal equilibrium with the SM plasma. Usually, when the temperature drops below the mass of these particles, also known as thermal bath particles, DM production becomes Boltzmann-suppressed, meaning the number density of the DM behaves as $n_{\chi}\propto \mathrm{exp}(-m_{\chi}/T)$  and the situation is known as freeze-in of the relic density. In this scenario, the produced dark matter is never abundant enough to annihilate among themselves and the coupling between the dark matter and the thermal plasma should be of order $\sim 10^{-11}$. Also, due to such feeble interactions, frozen-in DMs avoid most direct detection and collider search methods. Some X-ray and gamma-ray observatories provide useful information on indirect detection of FIMPs though \cite{Essig:2013goa}. If, in the near future, the experiments designed to provide proof of the WIMP paradigm don’t produce significant results, that avenue of thought will be under serious question. On the other hand, with the discovery of any evidence of WIMP-type dark matter, the FIMP scenario would be ruled out. Therefore, with such a win-win situation, the FIMP scenario is equally predictive as the WIMP one. 

On the other hand, Leptoquarks (LQs) \cite{Buchmuller:1986zs, Dorsner:2016wpm}, hypothetical particles with bosonic degrees of freedom, have become a very popular beyond Standard Model (BSM) candidate to address several experimental anomalies \cite{Bauer:2015knc,Bhattacharya:2016mcc,Crivellin:2017zlb,Hiller:2017bzc,Buttazzo:2017ixm,Chen:2017hir,Matsuzaki:2017bpp,Kumar:2018kmr,LHCb:2014vgu,Bifani:2017gyn,LHCb:2017avl,LHCb:2013ghj,BaBar:2013mob,Belle:2015qfa,LHCb:2015gmp,Muong-2:2006rrc}. These particles originate from quark-lepton unification theory \cite{Pati:1974yy, Georgi:1974sy} and thus can couple to both quarks and leptons and, as a result, can simultaneously be used as a solution to those anomalies. The WIMP paradigm corresponding to the Freeze-out scenario needs some portal through which DM can communicate with the SM particles and depending on the nature of the particles under consideration, the portal can be of different types. Such a portal has recently been explored in the context of scalar LQ (SLQ) and scalar DM in Ref.\cite{Choi:2018stw,DEramo:2020sqv,Sahoo:2021vug}. The same paradigm with SLQ and vector DM (VDM) cases has been studied in \cite{Baek:2012se,Sakaki:2013bfa,Baek:2014goa,Chao:2014ina,Chen:2015dea,Barman:2020ifq,Angelescu:2021lln}, whereas the fermionic DM (FDM) has been studied in \cite{Queiroz:2014pra,Mandal:2018czf}. On the contrary, although under the Standard Model (SM) gauge group a FIMP can be a new scalar singlet ~\cite{McDonald:2001vt,Yaguna:2011qn} or a fermion singlet \cite{Yaguna:2023kyu}, rendering interesting phenomenology, similar studies in the FIMP paradigm are not frequently available in the literature and this analysis is motivated to fill in that void.  

Therefore, in this article, we are specifically interested in exploring the FIMPs or DM freeze-in scenario in the presence of an SLQ and an SDM with a quartic coupling between them. We estimated the parameter spaces, including the masses and gauge couplings of DM with thermal bath particles to satisfy the freeze-in conditions as well as to produce the correct DM relic density. We have found that due to the smallness of such couplings, in the freeze-in scenario, heavy particles like LQs can easily be accommodated in the thermal bath and are perfectly allowed by all current experimental limits. For the same reason, the stability and perturbativity of the scalar potential don’t impose any significant constraint on our analysis as well.

This article is organised as follows: In Sec.~\ref{Sec:freeze-in with LQ}, we describe the basic framework considered for our analysis, including the Lagrangian and relic density generation. Sec.~\ref{sec:Results} consists of relevant details of the calculation, the results and subsequent discussions on the free parameters of this analysis and their constraints. Finally, we shall conclude in Sec.~\ref{Sec:conclusion}.


\section{Freeze-in scenario with leptoquarks} \label{Sec:freeze-in with LQ}

\subsection{The framework} \label{subsec:framework}

As described above, in the \textit{Freeze-in} scenario of dark matter (DM), the dark sector and the visible sector are thermally separated all the time. As a result, the initial dark matter density, which is considered to be negligible in the early universe, evolves with time. In this analysis, we extend the SM by a real singlet scalar $\chi$ which has direct interactions with both SLQ and the Higgs doublet, through quartic couplings. Stability of the dark matter is ensured by a $\mathbb{Z}_2$ symmetry under which only the dark matter changes as $\chi\rightarrow -\chi$, but other particles don't get affected. The production of DMs, as we will explain in detail in the following subsections, is strongly dependent on the electroweak symmetry breaking (EWSB), which gives a vacuum expectation value (VEV) to only the Higgs particles. Before the EWSB, the DM production is dominated by the scattering processes of LQs and Higgs, whereas afterwards the DMs are dominantly produced from the Higgs particle decays and Higgs-mediated decays of Leptoquarks.

 For LQs, it has been shown in Refs. \cite{Choi:2018stw, DEramo:2020sqv} that scalar $SU(2)_L$ singlet ($S_1$) and triplet ($S_3$) LQs which can be possible NP candidates to address the recent B-anomalies, can also act as a DM portal. Their $(SU(3)_c, SU(2)_L, U(1)_Y)$ quantum numbers are given by $(\bar{\mathbf{3}},1,1/3)$ and $(\bar{\mathbf{3}},3,1/3)$, respectively, following the conventions used in Ref.~\cite{Dorsner:2016wpm}. These LQs interact with matter particles through renormalizable Yukawa interactions with fermion currents carrying a net fermion number $F$, defined as $F = 3B + L$,
with $B$ and $L$ the baryon and lepton numbers, respectively. We consider the interactions with  $|F|=2$ which are given by \cite{Dorsner:2016wpm}

\bea 
\mathcal{L}_{S_1} & \supset & \big(y_1^{LL}\big)_{ij} \overline{Q}^{C,i,a}_L (i\sigma^2)^{ab}S_1 L_L^{j,b} +  \big(y_1^{RR}\big)_{ij} \overline{u}^{C,i}_R (i\sigma^2)^{ab}S_1 e_R^j \nnu
&&  + \big(y_1^{\overline{RR}}\big)_{ij} \overline{d}^{C,i}_R (i\sigma^2)^{ab}S_1 \nu_R^j
+ \big(\overline{z}_1^{RR}\big)_{ij} \overline{Q}^{C,i,a}_L (i\sigma^2)^{ab}S_1^* Q_L^{j,b} + h.c. \label{eq:S1-SM interxn} \\
\mathcal{L}_{S_3} & \supset & \big(y_3^{LL}\big)_{ij} \overline{Q}^{C,i,a}_L (i\sigma^2)^{ab}(\sigma^k S_3^k)^{bc}L_L^{j,c} + \big(\overline{z}_3^{LL}\big)_{ij} \overline{Q}^{C,i,a}_L (i\sigma^2)^{ab}\big[(\sigma^k S_3^k)^{\dagger}\big]^{bc}Q_L^{j,c} +  h.c., \label{eq:S3-SM interxn}
\eea
where $L,R$ and $C$ in the superscript represents the chirality and charge conjugation $(C)$ of the fermions respectively, $i,j=1,2,3$ are flavor indices, $a,b=1,2$ represents the $SU(2)$ indices $\sigma^k$ are the Pauli matrices and $S_3^k$ are components of $S_3$ (a triplet under $SU(2)$ space). $\big(y_i^{LL/RR}\big)_{ij}$ are arbitrary $3\times 3$ Yukawa coupling matrices describing the strength between LQ and quark-lepton pairs. On the other hand, the z couplings in the above equations represent an antisymmetric Yukawa coupling matrix in flavor space between LQ and quark-quark (diquark) pairs. These $z$ couplings potentially can lead to proton decay. Therefore, such interactions need to be avoided, and that can be done in several ways. One of the ways is to consider a discrete symmetry under which SM leptons and the LQs are assigned opposite parity or by some geometrical arguments (for details see Ref.~\cite{Aydemir:2019ynb} and the references therein). Thus we also consider a discrete symmetry, following Ref.~\cite{Bauer:2015knc}, to prevent such diquark interactions from appearing and choose scalar leptoquarks for our analysis, particularly focusing on $S_1$ for simplicity.

Therefore, the most general renormalizable Lagrangian for the above-mentioned framework takes the form 

\be \label{eq:LTotal}
\mathcal{L_{\rm Total}} = \mathcal{L_{\rm Higgs}} + \mathcal{L_{\rm LQ}} + \mathcal{L_{\rm SDM}}+ \mathcal{L_{\rm SDM}^{\rm Higgs}} + \mathcal{L_{\rm SDM}^{\rm LQ}} + \mathcal{L_{\rm LQ}^{\rm Higgs}} 
\ee
where 
\bea
\label{eq:LHiggs}
\mathcal{L_{\rm Higgs}} &=& - m_h^2 H^{\dagger}H  - 
\lambda_{H} |H^{\dagger}H|^2  \\
 \label{eq:LLQ}
\mathcal{L_{\rm LQ}} &=& |D_{\mu}S_{LQ}|^2 - m_{\rm LQ}^2 |S_{LQ}|^2 -  \frac{\lambda_{\Delta}}{4}|S_{LQ}|^4,\\
\label{eq:LSDM}
\mathcal{L_{\rm SDM}} &=& \frac{1}{2}(\partial_{\mu}\chi)^2 -\frac{1}{2} m_S^2\chi^2 -\frac{\lambda_{\chi}}{4} \chi^4, \\
\mathcal{L_{\rm SDM}^{\rm Higgs}} &=&  - \frac{\lamchiH}{2}
 H^{\dagger}H \chi^2, \label{eq:LSDMHiggs} \\
\mathcal{L_{\rm SDM}^{\rm LQ}} &=& - \frac{\lamchiD}{2} \chi^2 | S_{LQ}|^2 \label{eq:LLQSDM} \\
\mathcal{L_{\rm LQ}^{\rm Higgs}} &=& - \lamHD H^{\dagger}H | S_{LQ}|^2. \label{eq:LLQHiggs}
\eea
The covariant derivative $D_{\mu}$ is defined as 
\be \label{eq:Cov-dervtv}
D_{\mu} = \partial_{\mu} + ig_1 Y_{LQ} B_{\mu} + ig_2 T^j W_{\mu}^j + ig_3 T^K G_{\mu}^K
\ee
with $Y_{LQ}$ as the LQ hypercharge, $T^j (j=1-3)$ and $T^K (K=1-8)$ are $SU(2)_L$ and $SU(3)_c$ generators respectively.

After the EWSB (aEWSB), the Higgs fields will get the VEV $(v_H)$ and thus the interaction given in Eqs. \ref{eq:LSDMHiggs} and \ref{eq:LLQHiggs}, together with Eq. \ref{eq:LLQSDM}, will be most relevant and the scalar potential would take the form 
\be 
\label{eq:Scalr Pot aEWSB}
V_S\supset -\frac{1}{4}(\lamchiH\chi^2 + \lamHD| S_{LQ}|^2)(2v_H h+h^2) .
\ee


\subsection{Relic density generation}
\label{subsec:relic density}

Following the basic idea of the freeze-in scenario, the initial DM number density was negligible and it was gradually produced from scattering and decays of the visible sector particles. Here we should describe the general processes following the Big Bang cosmology and radiation-dominated universe. The details of those processes are discussed in detail in \ref{Subsec:Freeze-in processes}. 

The production of DM $(\chi)$ happens in two phases. bEWSB, the dominant production channels are $\lamchiH\chi^{\dagger}\chi H^{\dagger}H $ and $\lamchiD\chi^{\dagger}\chi| S_{LQ}|^2$ interaction terms, leading to Feynman diagrams shown in Fig. \ref{Fig:H LQ annihln to DM}. aEWSB, the DM production is dominated by $h\rightarrow\chi^{\dagger}\chi$ for the condition $\mchi < m_h/2$, although similar decays from LQs are also possible. It is convenient to express any early stage scenario of the universe as a function of the SM temperature, $T$. Therefore, for the Higgs, we define $\langle\sigma v\rangle^T_H= \langle\sigma v\rangle_{HH\rightarrow \chi\chi}\Theta(T-T_{\rm EW})+\langle\Gamma_{h\rightarrow\chi\chi}\rangle\Theta(T_{\rm EW}-T)$, where $T_{\rm EW}\sim 160\GeV$ represents the electroweak phase transition temperature (EWPT) \cite{Kajantie:1995kf}.  

In the freeze-in scenario, the DMs are considered to be out of thermal bath, which means the number of DM particles in the hidden sector should not be such that they will start annihilating back to the SM particles. Therefore, the condition 
\be \label{eq:DM freeze-in condn}
\frac{n_{j}^{\text eq}\langle \sigma v_{M\slashed{o}l}\rangle ^T_{jj\rightarrow \chi\chi}}{\mathcal{H}} < 1,
\ee 
,needs to be satisfied. In Eq.~\ref{eq:DM freeze-in condn} the quantity $\left\langle \sigma v_{M\slashed{o}l}\right\rangle$, the thermal average of the annihilation cross-section of bath particles $(j)$ to DM particles, times the Moller velocity, provides a very important quantity for the calculation of the number density of DM at thermal equilibrium. For the FIMP, it is defined as \cite{Gondolo:1990dk}
\be \label{eq:Thermal avg sigmav defn}
\langle\sigma v_{M\slashed{o}l}\rangle_{jj\rightarrow \chi\chi} = \frac{1}{8 m_j^4 T K_2^2[\frac{m_j}{T}]}\int_{4m_j}^{\infty} ds (s - 4m_j^2) \sqrt{s}  \sigma_{jj\rightarrow \chi\chi}(s)K_1\big(\frac{m_j}{T}\big)
\ee
where $K_i$s are the modified Bessel functions of order $i$. $\mathcal{H}$ represents the Hubble parameter relating the total energy density of the universe $\rho$ and the Planck Mass $(M_{\rm Pl})$ as $\mathcal{H}=\sqrt{\frac{8}{3}\pi\rho}/M_{\text Pl}$ and $n^{\text eq}_j$ is the number density of the decaying particle $j$ at thermal equilibrium, defined as
\bea 
n^{\text eq}_j (t) &=& g_j (T) \int \frac{d^3p}{(2\pi)^3}f(p,t)\label{eq:neq defn}\\
\rho (t) &=& g_j (T) \int \frac{d^3p}{(2\pi)^3}f(p,t) E , \label{eq:rho defn}
\eea
with $g_j$ being the internal degrees of freedom of the corresponding particle, $f(p,t)$ is the distribution function of momentum $(p)$ and time $(t)$ of the early universe and the energy is given by $E = \sqrt{p^2+m_{\chi}^2}$.
Therefore, for heavy particles like LQs to be in the thermal bath, in the freeze-in scenario, the condition 
\be 
\label{eq:LQ freeze-in condn}
\frac{n_{\Delta}^{\text eq} \left\langle \sigma  v_{\Delta\Delta\rightarrow \chi\chi}\right\rangle}{\mathcal{H}} < 1
\ee
must be satisfied.

For annihilation of two particles, 1 and 2 with initial number densities given by $n_i (i=1,2)$, into two others, 3 and 4,
the Boltzmann equation describing the number density is given by \cite{Gondolo:1990dk}
\be \label{eq:Boltzmann eqn 1}
\frac{dn_i}{dt}+3 \mathcal{H} n_i = \langle\sigma v\rangle\big(n_in_j - n^{\text eq}_in^{\text eq}_j\big)\quad\quad (i,j=1,2~\text and~ i\neq j). 
\ee
where $n^{\text eq}_{i,j}$ are the particle number densities generated when they are at thermal equilibrium. When 1 and 2 are identical particles, $n=n_1=n_2$ and the above equation can be written as
\be \label{eq:Boltzmann eqn 2}
\frac{dn}{dt}+3 \mathcal{H} n = \langle\sigma v\rangle\big[n^2 - (n^{\text eq})^2\big]. 
\ee
 Similarly, for the decay of Higgs to DM particles, the Boltzmann equation for the  evolution of $n_{\chi}$ is given by~\cite{Bernal:2017kxu},
 \be \label{eq:Boltzmann eqn 3}
\frac{dn_{\chi}}{dt}+3 \mathcal{H} n_{\chi} = 2\Gamma_{h\to \chi\chi}\frac{K_1(m_h/T)}{K_2(m_h/T)}\big[n^2 - (n^{\text eq}_h)^2\big]. 
\ee
where $K_j$ are modified Bessel functions of the second kind, $\Gamma_{h\to \chi\chi}$ is the decay width and $n^{\text eq}_h$
is the equilibrium number density of $h$.
Since in the freeze-in scenario, DM is considered to be in the hidden sector, not in thermal equilibrium with visible sector particles, its distribution function $f(p,t)$ is zero and thus, following Eq.~\ref{eq:neq defn}, the first term in Eqs.~\ref{eq:Boltzmann eqn 2} and~\ref{eq:Boltzmann eqn 3} can be neglected. Therefore, the Boltzmann equation describing the dark matter number density from thermal bath particle interactions is given by,
\be \label{eq:Boltzmann eqn freeze-in-1}
\frac{dn_{\chi}}{dt}+3 \mathcal{H} n_{\chi} = - (n^{\rm eq})^2\big[\big(\langle\sigma v\rangle_{\Delta\Delta\rightarrow \chi\chi}+\langle\sigma v\rangle_{HH\rightarrow \chi\chi}\big)\Theta(T-T_{\rm EW})+\langle\Gamma_{h\rightarrow\chi\chi}\rangle\Theta(T_{\rm EW}-T)\big]. 
\ee 
\noindent
Using the expression for the yield, $ Y = n_{\chi}/s $, the relation between the number of DM particles $(n)$ and the total entropy density of the universe $(s)$, we can rewrite Eq. \ref{eq:Boltzmann eqn freeze-in-1} as  

\be \label{eq:Boltzmann eqn freeze-in-2}
\frac{dY_{\chi}}{dt} = -s\langle\sigma v\rangle_{\text {All processes}}\big(Y_{\chi}^2 - (Y^{\rm eq})^2\big), 
\ee  
where $\langle\sigma v\rangle_{\text {All processes}}$ includes all DM producing processes such as $HH\rightarrow \chi\chi, \Delta\Delta\rightarrow \chi\chi$ and $h\rightarrow \chi\chi$ and 
\be \label{eq:entropy density reln}
s = h_{\rm eff} \frac{2\pi^2}{45}T^3,
\ee
$h_{\rm eff}$ is the effective number of degrees of freedom of the bath particle.

Since entropy is a function of temperature $(T)$, we can express Eq.\ref{eq:Boltzmann eqn freeze-in-2} as 
\be 
\frac{dY_{\chi}}{dz} = \frac{s}{\mathcal{H} z}\langle\sigma v\rangle_{\text {All processes}}(Y_{\chi}^{\rm eq})^2,
\label{eq:Boltzmann eqn freeze-in-3}
\ee 
where $z = M_{sc}/T$, with $M_{sc}$ being some high-energy scale and $T$ being the temperature. $Y_{\chi}^{\rm eq}$ represents the DM yields at equilibrium from all possible processes. Inserting Eq.\ref{eq:Thermal avg sigmav defn} in Eq.\ref{eq:Boltzmann eqn freeze-in-3} and expressing $T$ as a function of $z$, we can perform the double integration to find the number density of DM particles through this relation
\bea \label{eq:Boltzmann eqn freeze-in-4}
Y_{\chi} &=& \int_{z_{\rm min}}^{z_{max}}dz\big(\frac{s}{\mathcal{H} z}\big)\big[ (Y_H^{\rm eq})^2\langle\sigma v\rangle_{HH\rightarrow \chi\chi}\Theta(T-T_{\rm EW})\nonumber \\
&+&\big((Y_{\Delta}^{\rm eq})^2\langle\sigma v\rangle_{\Delta\Delta\rightarrow \chi\chi}+(Y_h^{\rm eq})\langle\Gamma_{h\rightarrow\chi\chi}\rangle\big)\Theta(T_{\rm EW}-T)\big]. 
\eea 
Therefore, the relic density of $\chi$ is approximately given by 
\be \label{eq:Relic density Defn}
\Omega h_{\chi}^2 \simeq 2.742\times 10^{8}m_{\chi}Y_{\chi} . 
\ee


\subsection {Dark matter production in freeze-in mechanism} \label{Subsec:Freeze-in processes}

Among several other possible ways to produce DM particles from the bath particles, one possibility is the scalar LQ acquiring the VEV and thus decaying to DM particles after the electroweak symmetry breaking. Although it is not uncommon to provide a VEV to the LQs in different phenomenological scenarios, it is not suitable for the present purpose due to several model-building complications that might arise. These complications include the breaking of $SU(3)$ color symmetry or introducing dangerous operators responsible for proton decays. Prevention of such operators is possible by the $Z_2$ symmetry, whereas we are assuming $Z_3$ symmetry for our case. Therefore, to avoid such issues, the LQ is not provided with the VEV and thus the production channel of the DM from the direct decay of the LQ is prohibited for the present purpose.

Consequently, following the basic features of the freeze-in mechanism in standard cosmology, the simplest scenarios of freeze-in dark matter production are dominated by $2 \rightarrow 2$ annihilation and $1 \rightarrow 2$ decay of visible sector particles and $LQs$ into DM \cite{Hall:2009bx}.

\paragraph{Freeze-in by $2 \rightarrow 2$ scattering of bath particles:}

One of the dominant methods of producing relic abundance in the freeze-in scenario is $2 \rightarrow 2$ scattering, where the DM ($\chi$), a scalar, interacts with other bath particles via the operators given in Eqs. \ref{eq:LSDMHiggs} and \ref{eq:LLQSDM} as shown in Fig.~\ref{Fig:H LQ annihln to DM}. The temperature in these cases must be greater than the reheat temperature $(T_{\rm RH})$ on which the dark matter abundance depends. The annihilation cross-sections for the relevant processes are given below:

\begin{enumerate}

\item $HH\rightarrow \chi\chi$, with $H$ being the SM Higgs field before the EWSB,

\be \label{eq:hhtoDM}
\sigma_{H H^* \rightarrow \chi\chi} = \frac{\lamchiH^2 }{64 \pi  s}\sqrt{\frac{ \left(s-4 m_{\chi} ^2\right)}{ \left(s-4m_h^2\right)}}
\ee

\item $\Delta\Delta \rightarrow \chi\chi$, Similar to the Higgs, Leptoquarks can also annihilate to DM particles bEWSB, as shown in Fig.~\ref{Fig:H LQ annihln to DM}, with cross-section given by~\cite{DEramo:2020sqv},
\be \label{eq:LQLQtoDM bEWSB}
\sigma_{\Delta\Delta \rightarrow \chi\chi} = \frac{N_c (2 T_{\Delta} + 1)\lamchiD^2}{16 \pi  s\left(s-m_h^2\right)^2}\sqrt{\frac{s-4 m_{\chi}^2}{s-4 m_{\Delta}^2}}.
\ee
In the above formula $N_c=3$ denotes the number of colors and $2 T_{\Delta} + 1$ stands for the LQ weak-isospin multiplicity. $T_{\Delta}=0$ represents a weak singlet LQ. Similarly, $ T_{\Delta} = 1/2,0$ represent a doublet and a triplet LQ respectively.

\begin{figure}[ht!] 
\centering
\begin{subfigure}{0.42\textwidth}
\begin{tikzpicture}
\begin{feynman}
\vertex (a) at (0, 0);
\vertex (b) at (-1, 1) {\(H\)};
\vertex (c) at (-1, -1) {\(H^{\dagger}\)};
\vertex (d) at (1, 1) {\(\chi\)};
\vertex (e) at (1, -1) {\(\chi\)};
\diagram* {
(b) -- [scalar] (a) -- [scalar] (c),
(d) -- [scalar] (a) -- [scalar] (e),
};
\end{feynman}
\end{tikzpicture}
\end{subfigure}
\quad\quad
\begin{subfigure}{0.42\textwidth}
\begin{tikzpicture}
\begin{feynman}
\vertex (a) at (0, 0);
\vertex (b) at (-1, 1) {\(\Delta\)};
\vertex (c) at (-1, -1) {\(\Delta\)};
\vertex (d) at (1, 1) {\(\chi\)};
\vertex (e) at (1, -1) {\(\chi\)};
\diagram* {
(b) -- [scalar] (a) -- [scalar] (c),
(d) -- [scalar] (a) -- [scalar] (e),
};
\end{feynman}
\end{tikzpicture}
\end{subfigure}
\caption{Tree-level Feynman diagrams, bEWSB, for annihilation of Higgs and LQ to real scalar dark matter.}
\label{Fig:H LQ annihln to DM}
\end{figure}

\end{enumerate}

\begin{figure}[H]
\centering
   \begin{tikzpicture}
\begin{feynman}
\vertex (a) at (0, 0);
\vertex (b) at (-2, 1) {\(\Delta\)};
\vertex (c) at (-2, -1) {\(\bar{\Delta}\)};
\vertex (d) at (2, 0);
\vertex (e) at (4, 1) {\(\chi\)};
\vertex (f) at (4, -1) {\(\chi\)};
\diagram* {
(b) -- [scalar] (a) -- [scalar] (c),
(a) -- [scalar, edge label'=\(h\)] (d),
(e) -- [scalar] (d) -- [scalar] (f),
};
\end{feynman}
\end{tikzpicture} 
\caption{Higgs-mediated tree-level Feynman diagram for annihilation of scalar LQs to scalar dark matter, aEWSB. }    
\label{Fig:LQ annihiln to DM aEWSB}
\end{figure}

After the EWSB, a pair of LQs can also annihilate to produce a pair of DMs via Higgs mediation, following Fig.~\ref{Fig:LQ annihiln to DM aEWSB} and the corresponding cross-section is given by

\be \label{eq:LQLQtoDM aEWSB}
\sigma_{\Delta\Delta \rightarrow \chi\chi} = \frac{N_c (2 T_{\Delta} + 1)v^4\lamchiH^2\lamHD^2 }{16 \pi  s\left(s-m_h^2\right)^2}\sqrt{\frac{s-4 m_{\chi}^2}{s-4 m_{\Delta}^2}}.
\ee

Other similar possible annihilation processes via Higgs mediation, with their cross-sections, are given in Appendix~\ref{appendix:other xscns}. Compared to the scattering processes of scalar particles, the same for vector particles are much suppressed due to the presence of a heavy mediator particle and the corresponding coupling strengths and therefore those processes are ignored in the present analysis.

\paragraph{Freeze-in by decay of bath particles:}

The most dominant production channel of the dark matter in the freeze-in scenario comes from the decays of bath particles. This happens below the reheating temperature ($T_{\rm RH}$), where the particles that freeze-in, being unstable, just decay to dark matter. In our case, leptoquarks don't get vacuum expectation values (VEVs) after the EWSB and thus don't decay to dark matter particles, whereas the abundance of the dark matter via the freeze-in mechanism in our setup would be through the Higgs particles $(h)$ decay after the EWSB. When DM is lighter than half the Higgs mass, the relevant process is shown in Fig.~\ref{Fig:Higgs decay to DM} 
\begin{figure}[H] 
\begin{center}
\begin{tikzpicture}
\begin{feynman}
\vertex (a) at (0, 0);
\vertex (b) at (2, 0);
\vertex (c) at (4, 1) {\(\chi\)};
\vertex (d) at (4, -1) {\(\chi\)};
\diagram* {
(a) -- [scalar, edge label'=\(h\)] (b),
(c) -- [scalar] (b) -- [scalar] (d),
};
\end{feynman}
\end{tikzpicture}
\end{center}
\caption{Decay of Higgs to DM particles}
\label{Fig:Higgs decay to DM}
\end{figure}
and the corresponding cross-section for the process is given by,

\be \label{eq:htoDMDM}
\Gamma_{h\rightarrow \chi\chi} = \frac{\lamchiH^2 v^2}{32 \pi m_h^2 }\sqrt{m_h^2-4 m_{\chi}^2}\,.
\ee


\section{Results and discussions}
\label{sec:Results}

As discussed above, this analysis has several free parameters, but not all of them are equally important for the DM phenomenology. Among these free parameters, self-couplings of Eqs.~\ref{eq:LHiggs},~\ref{eq:LLQ} and \ref{eq:LSDM} are important for the stability of the scalar potential and thermal correction of masses. Other free parameters include the LQ-Higgs coupling ($\lamHD$), two DM couplings ($\lamchiH, \lamchiD$), LQ mass $(m_{\Delta})$ and DM mass $(m_{\chi})$. Out of these, DM couplings and mass contribute significantly towards the DM phenomenology and hence exploring these parameter spaces will be detailed in this section.

\textbullet~\textbf{Calculation details}: To estimate the constraints on the relevant couplings, the first step was to use the cross-section formulae mentioned in equations~\ref{eq:hhtoDM}- \ref{eq:htoDMDM}. Then a numerical integration was performed in \textit{Mathematica}\cite{Mathematica}, following Eqn.\ref{eq:Thermal avg sigmav defn}, over the range of $s$ to calculate the thermal average of the cross-section. In those integrations, the lower limits were always the pole masses of the annihilating particles and for the upper limit an arbitrarily high mass value $(\sim 10^{10})$ was chosen. Number density calculations involve the internal degrees of freedom (d.o.f) of the corresponding particle, and those change during the temperature evolution. For the Higgs, these d.o.f were taken as 4 and 1 before and after the EWSB, respectively. For the LQ, the d.o.f was calculated using the formula $N_c(2T_{\rm LQ}+1)$, where $N_c$ and $T_{\rm LQ}$ represent the color and weak isospin factor of LQ, respectively. Since the electroweak transition is an important temperature-dependent phenomenon which acts as a limiting case, special care was taken to generate the plots highlighting this transition region. As explained in sub-section \ref{subsec:relic density}, to generate the final DM yield, Boltzmann's differential equation, given in Eqn.~\ref{eq:Boltzmann eqn freeze-in-4}, was solved numerically within a typical range of $10^{-6} - 10^4$ for the variable $z$. In this regard, it is worth mentioning that the value of the effective number of entropic degrees of freedom $h_{\rm eff}$, given in Eqn.~\ref{eq:entropy density reln}, was chosen to be $106.75$ for the electroweak epoch condition $(T > 100\text{ GeV})$ in the very early universe. Although it is generally not a standard procedure to assume a fixed $h_{\rm eff}$ when calculating the DM yield in a freeze-in scenario, because freeze-in occurs over a wide range of temperature, it was chosen for a couple of reasons. First, a renormalizable interaction between the scalar DM and the scalar LQ was considered and thus the calculation was stable in the infrared (IR) region $(T>m_{\rm LQ})$. Since the bulk of DM production occurs at this high-temperature region where all Standard Model degrees of freedom are fully active, evaluating \(h_{\mathrm{e}ff}\) at a constant value of \(106.75\) introduces a minor numerical error which does not impact our qualitative conclusions. Second, the purpose of this analysis was to estimate an order-of-magnitude DM yield rather than a very precise calculation. Finally, the relic density was calculated following Eqn.~\ref{eq:Relic density Defn}.

\begin{figure}[htbp]
\captionsetup{justification=raggedright,singlelinecheck=off} 
\centering
\includegraphics[width=12 cm]{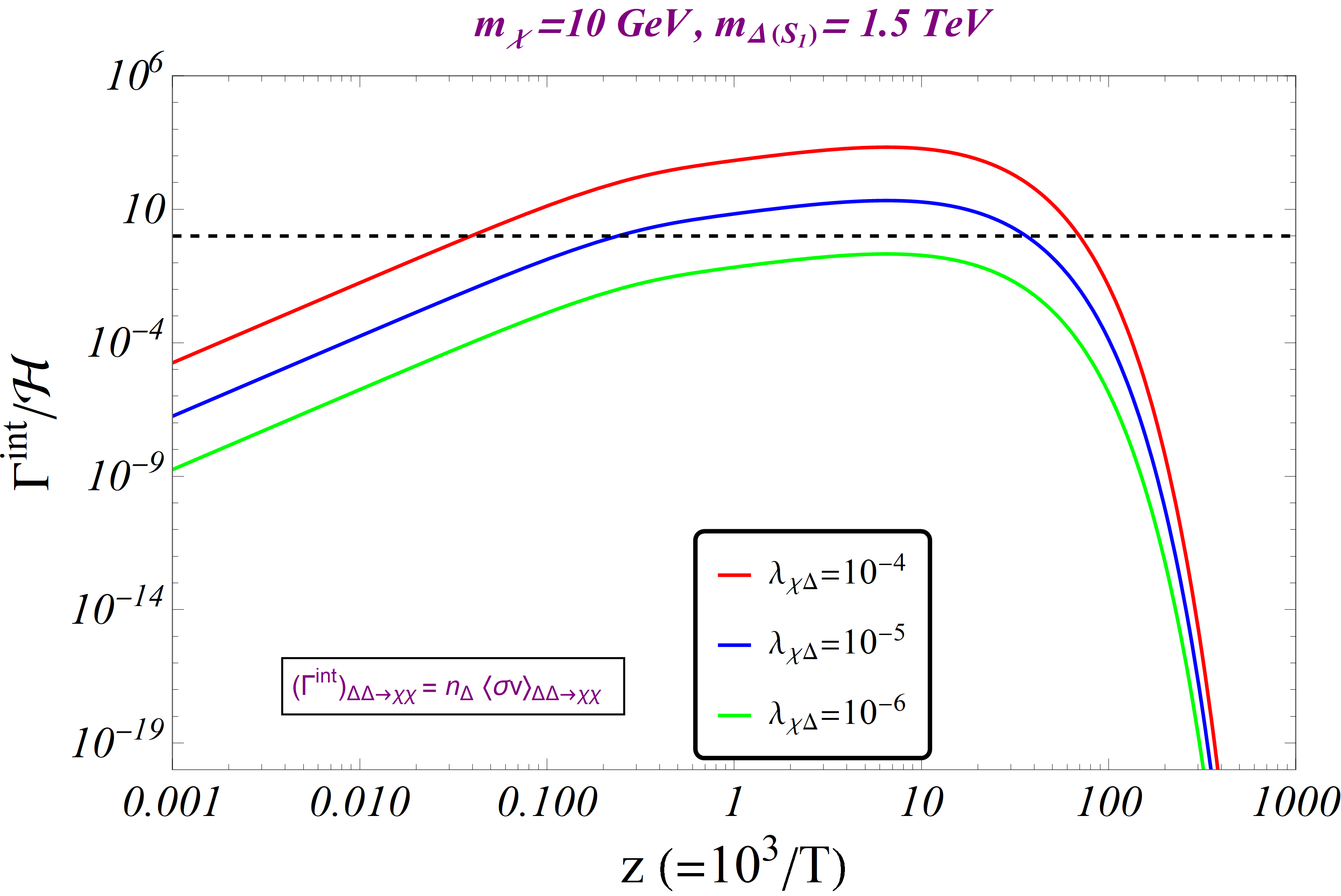}
\caption{The range of DM-LQ coupling $(\lamchiD)$ needed to satisfy the freeze-in condition (Eq.~\ref{eq:LQ freeze-in condn}), for DM and LQ mass $10 \GeV$ and $1.5 \TeV$ respectively, is shown in the figure. Different colors represent different coupling strengths as mentioned in the legend. The horizontal dashed line corresponds to the situation where the interaction rate of singlet LQ scattering to DM particles $(\Gamma^{\rm int})$ equals the Hubble parameter $(\mathcal{H})$ ($n_{\text eq} \left\langle \sigma  v\right\rangle = \mathcal{H}$).}
\label{Fig:LQ-DM coupling range}
\end{figure}

\textbullet~\textbf{Constraint on $\lamchiD$:} In the freeze-in scenario, the DM and bath particles remain decoupled from each other. Since the production of the DM in the scalar sector mostly happens from the Higgs and LQs by scattering processes (as shown in Fig.~\ref{Fig:H LQ annihln to DM}), we estimate the bounds on these two DM couplings $(\lamchiH,\lamchiD)$ from the freeze-in conditions given by Eqs.~\ref{eq:DM freeze-in condn} and \ref{eq:LQ freeze-in condn}. Particularly, the coupling $\lamchiD$ is relevant for the scattering of LQs directly to DM particles, both before and after EWSB. Therefore, we receive a constraint on it from the process $\Delta\Delta\to\chi\chi$, as shown in Fig.~\ref{Fig:LQ-DM coupling range}. For obvious reasons, the upper limit obtained from the time bEWSB will remain the same as the aEWSB.

\begin{figure}[htbp] 
\captionsetup{justification=raggedright,singlelinecheck=off}
\centering
\includegraphics[width=12 cm]{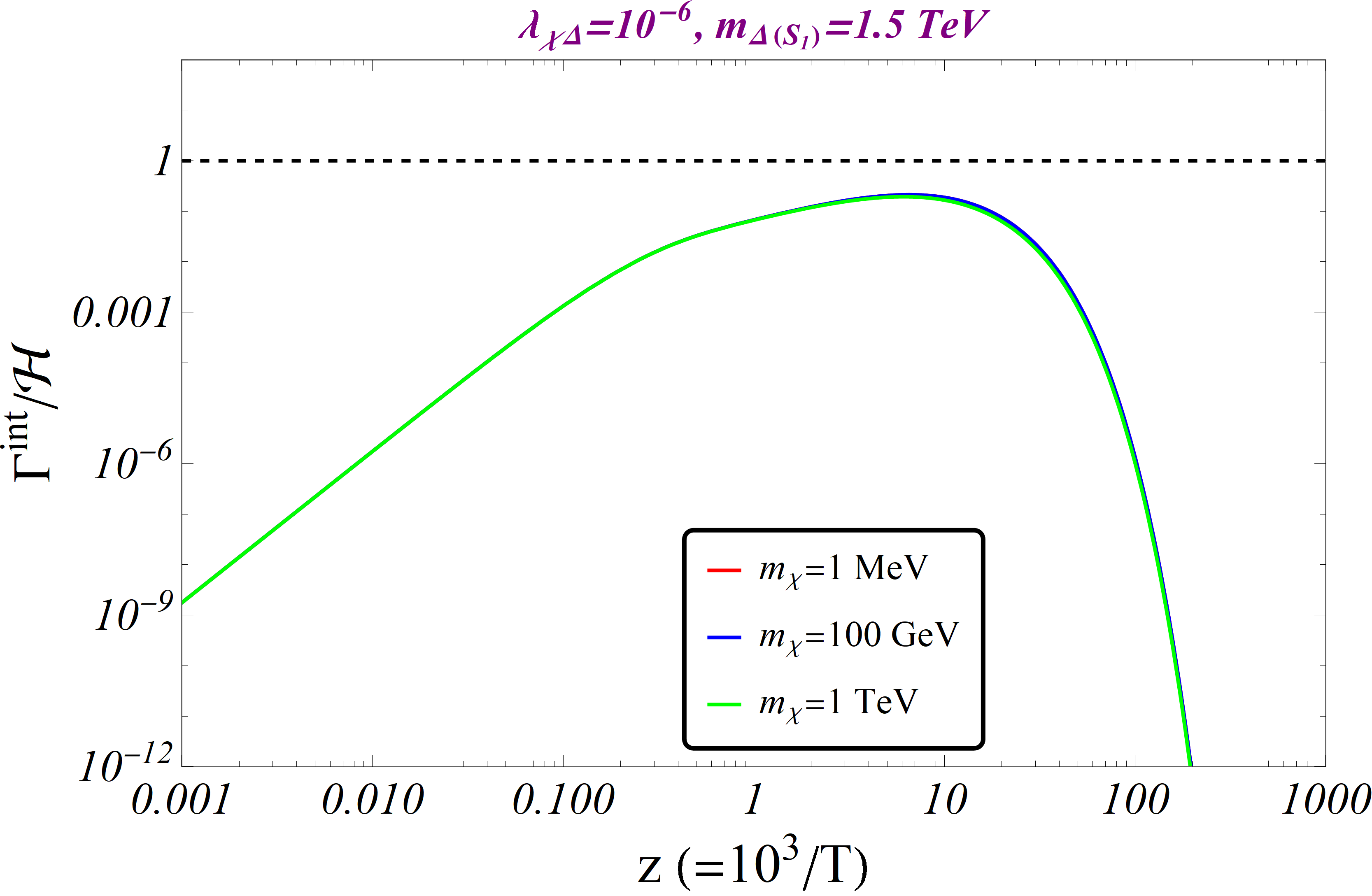}
\caption{The ratio of interaction rate $(\Gamma^{\rm int})$ of the singlet LQ and the Hubble parameter $(\mathcal{H})$ as a function of the temperature for different values of DM mass ($m_{\chi}$) with the DM-LQ coupling being fixed at $(\lamchiD =) 10^{-6}$.}
\label{Fig:DM prodcn for diff DM masses}
\end{figure}

Fig.~\ref{Fig:LQ-DM coupling range} depicts the ratio of the interaction rate of singlet LQs annihilating to DM particles $(\Gamma^{\rm int})$ to the Hubble parameter $(\mathcal{H})$ against the quantity $z$, which is defined previously in Eq.~\ref{eq:Boltzmann eqn freeze-in-3}. Green, blue and red colors represent the rates of annihilation of LQs to the DM particles for the DM-LQ coupling ($\lamchiD$) values of $10^{-4}, 10^{-5}$ and $10^{-6}$, respectively. Clearly, from this figure we find that for an arbitrarily chosen DM mass of $10 \GeV$ and a carefully chosen LQ mass of $1.5 \TeV$, $\lamchiD \lesssim 10^{-6}$ satisfies the freeze-in condition.  Now, it would be relevant to examine whether this coupling value is also acceptable for other DM masses as well. Following Eq.~\ref{eq:LQLQtoDM bEWSB}, it is understood that the DM production rate would not be affected significantly due to the presence of a heavier LQ mass in the denominator. This is indeed shown in Fig.~\ref{Fig:DM prodcn for diff DM masses} where the interaction rate is observed for different DM masses, ranging from $\MeV$ to $\TeV$, but for the fixed coupling $\lamchiD = 10^{-6}$. Clearly, it shows the independent nature of this coupling for varying DM masses. 

\begin{figure}[H]
\captionsetup{justification=raggedright,singlelinecheck=off} 
\centering
 \includegraphics[width=12 cm]{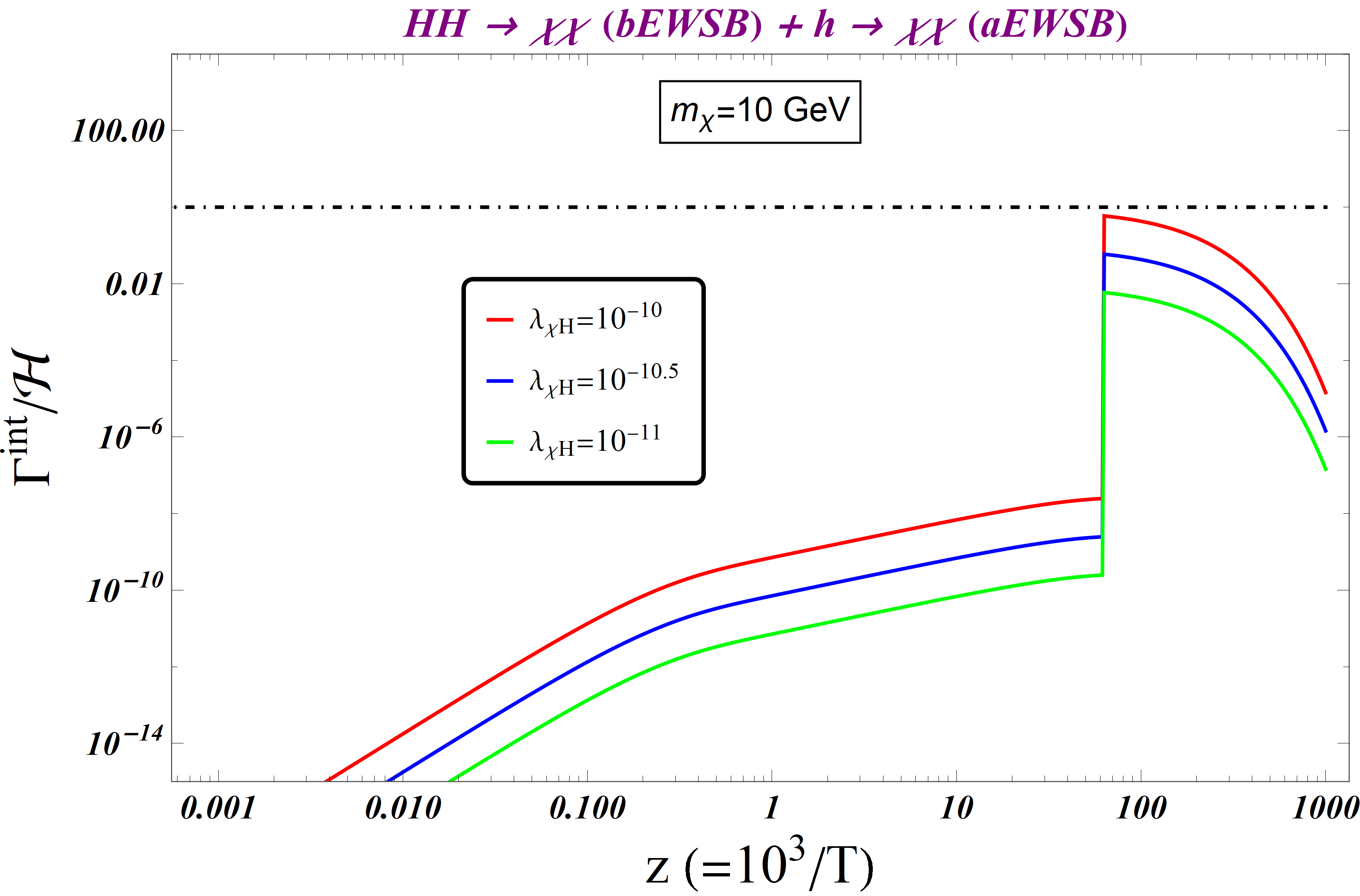} 
 \caption{The plot shows the range of DM-Higgs coupling $(\lamchiH)$ obtained from the freeze-in condition. Before the EWSB, the Higgs fields $(H)$ are annihilated to DM particles ($HH\to\chi\chi$) and after the EWSB, when the Higgs particles $(h)$ are getting VEVs, they are decaying to DM particles ($h\to\chi\chi$). Three different colors present three different $(\lamchiH)$ values as mentioned in the legend.}
\label{Fig:Higgs coupling bEWSB aEWSB}
\end{figure}

\begin{figure}[b]
\captionsetup{justification=raggedright,singlelinecheck=off}
\centering
\includegraphics[width=12 cm]{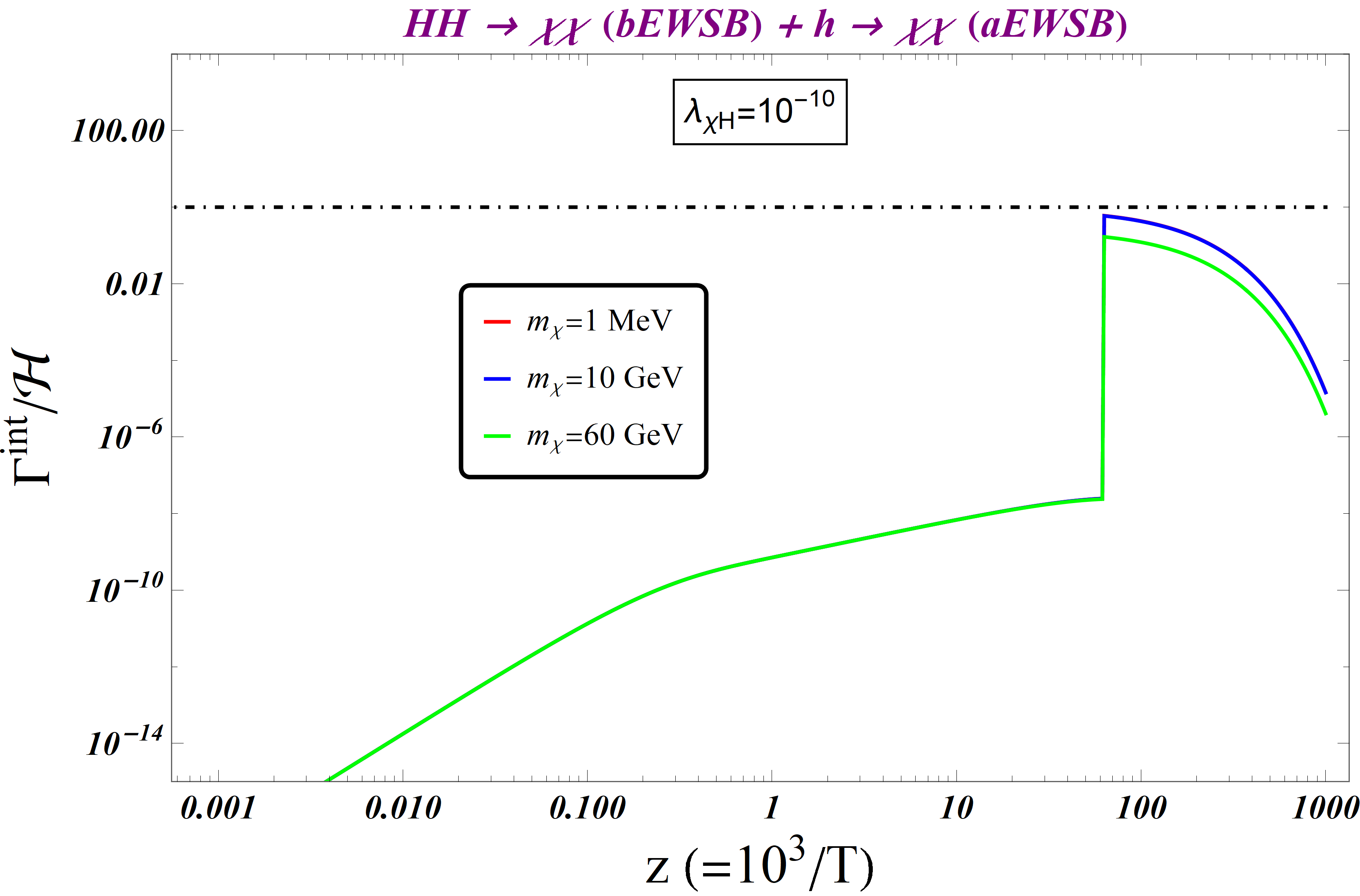} 
\label{Fig:LQ DM yield with diff DM masses}
\caption{The plot shows how the total DM production rate from the Higgs particle are changing with the change of DM mass $\mchi$ (from $\MeV$ to $\GeV$ range), for a fixed coupling $\lamchiH=10^{-10}$. As mentioned before, the horizontal dashed line corresponds to the situation ($n_{\text eq} \left\langle \sigma  v\right\rangle = \mathcal{H}$).}
\label{Fig:Yield for same DM-H coupling diff DM mass}
\end{figure}

\textbullet~\textbf{Constraint on $\lamchiH$:} The value of this coupling for which the freeze-in condition is satisfied is shown in Fig.~\ref{Fig:Higgs coupling bEWSB aEWSB}. Before the EWSB, when the DM particles are created by the Higgs field $(H)$ annihilation, $\lamchiH \lesssim 10^{-6}$ satisfies the freeze-in condition. On the other hand, after the EWSB, the dominant contribution to the DM production comes from the Higgs particle $(h)$ decay and $\lamchiH \lesssim 10^{-10}$ satisfies the same condition in this scenario. Therefore, it was prudent at first to consider different $\lamchiH$ values for an arbitrary DM mass (which we chose as $10~\GeV$) to satisfy the freeze-in condition. This is shown in Fig.~\ref{Fig:Higgs coupling bEWSB aEWSB}. Next, it would be natural to ask how 
this coupling behaves to satisfy the freeze-in condition, for different DM masses. This is shown in Fig.~\ref{Fig:Yield for same DM-H coupling diff DM mass}, where the interaction rate for three different mass values of $1 \MeV, 10 \GeV$ and $60 \GeV$, for the coupling strength $\lamchiH = 10^{-10}$, are presented in red, blue and green colors respectively. We find that for all these masses, the ratio of $(\Gamma^{\rm int}/\mathcal{H})$ lies below 1 (horizontal dashed line) and thus satisfies the required condition. Therefore, $\lamchiH \lesssim 10^{-10.5}$ is chosen to be the accepted value for this coupling in the present analysis. 

The DM couplings, $\lamchiH$ and $\lamchiD$, receive bounds from direct detection experiments such as XENON1T \cite{xenon1t}, PANDA \cite{panda}, LUX \cite{LUX:2016ggv}, etc., and it is customary for any dark matter study to obey the bounds obtained from these experiments. Although in this regard it must be stressed that these bounds apply only to the WIMP regime of the singlet model and no comparable bound exists in the freeze-in regime. Because of the tiny couplings of DM to bath particles, as is evident from our analysis also, it is easier to evade such constraints in the freeze-in scenario. 
The $\lamchiH$ coupling also receives a limit from the invisible decay width of the Higgs experiment \cite{CMS:2018yfx}, and it is applicable when the DM mass range is $\leq m_h/2$ and Eq.~\ref{eq:htoDMDM} is valid. With the observed (expected) upper limit of 0.33 (0.25) on the branching fraction of the Higgs boson decay to invisible particles $(\mathcal{B}(H \rightarrow inv))$ at $95\%$ CL, from Eq.~\ref{eq:htoDMDM}, we find that the constraint is well satisfied in our setup.

 
\textbullet~\textbf{Constraint on $\lamHD$:} From the scalar potential, we find that the coupling $\lamHD$ is not responsible for the DM production before the EWSB, but this becomes relevant for the DM production from annihilation of LQs after the EWSB (see Fig.~\ref{Fig:LQ annihiln to DM aEWSB}). This coupling is also relevant for di-Higgs production from LQs~\cite{Enkhbat:2013oba,DaRold:2021pgn} where it is usually considered to be of order unity. Therefore, mostly in this analysis it is considered as $\lamHD = 1$. This quartic Higgs-LQ coupling doesn't have any known bound reported from LHC or any other experiment.


\textbullet~\textbf{Constraint on $m_{\Delta}$:}  Constraints on LQ masses can arise from several interactions with matter particles which propagate through the Yukawa couplings. These interactions include Kaon, $B$ and $D$ meson decays, $(g-2)$ of charged leptons, neutral meson mixing, electric dipole moments etc. Collider experiments provide lower limits on the LQ mass via cross-section of pair or single production modes.  Assuming the third generation of leptoquarks and considering different decay modes, such limits can be set with $95\%$ confidence level in the range $0.450 - 1.88 \TeV$~\cite{CMS:2023qdw}, but with a significantly large coupling of $2.5$. For a summary on LQ mass limits and corresponding experimental searches, see Ref. \cite{PDG:2024cfk} and references therein. 

\begin{figure}[H]
\centering
\begin{subfigure}{0.4\textwidth}
\centering
\includegraphics[width=1.1\linewidth]{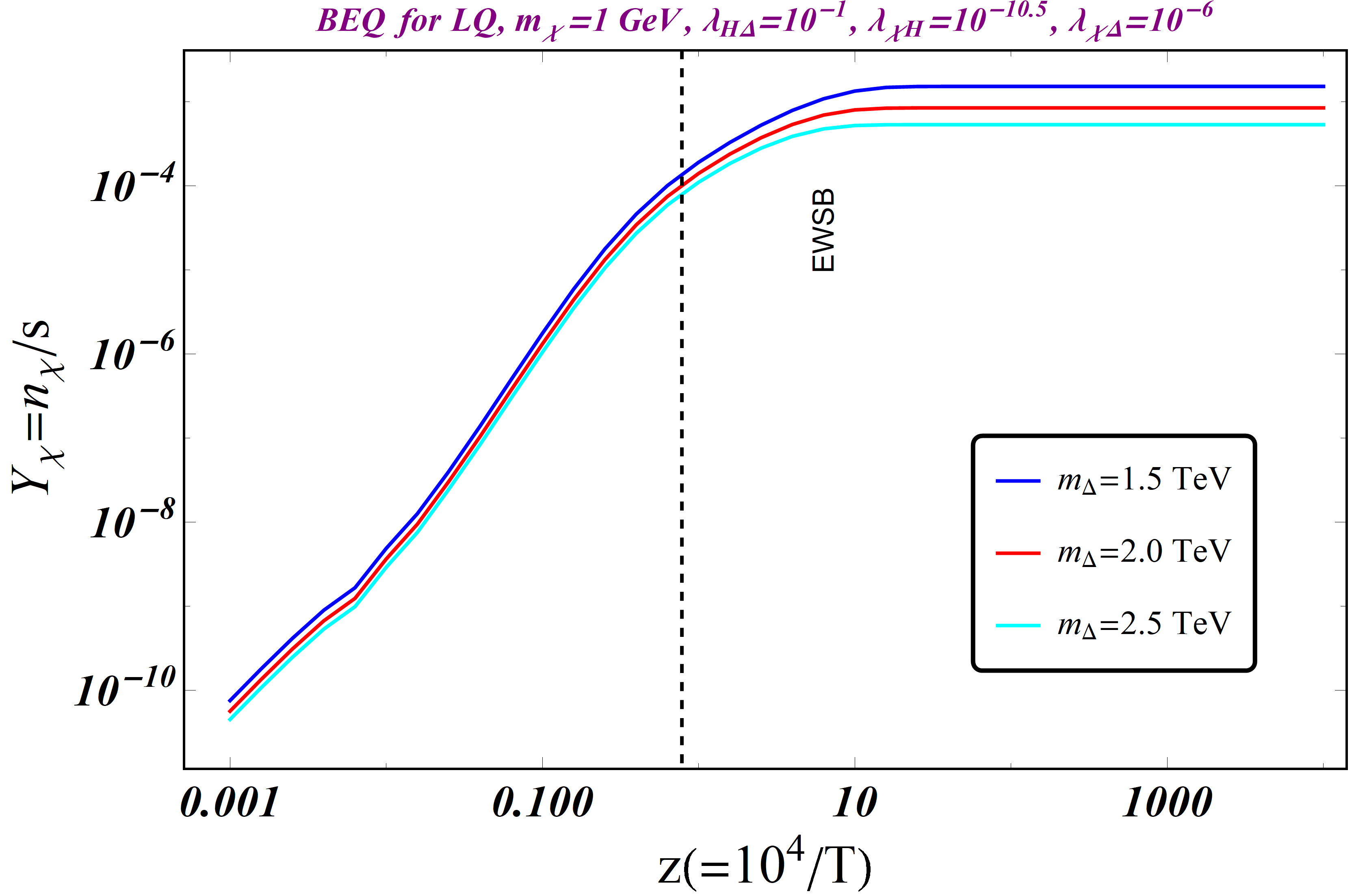} 
\caption{DM yield dependence on LQ mass}
\label{Fig:LQ DM yield with diff LQ masses}
\end{subfigure}
\quad\quad
\begin{subfigure}{0.4\textwidth}
\centering
\includegraphics[width=1.1\linewidth]{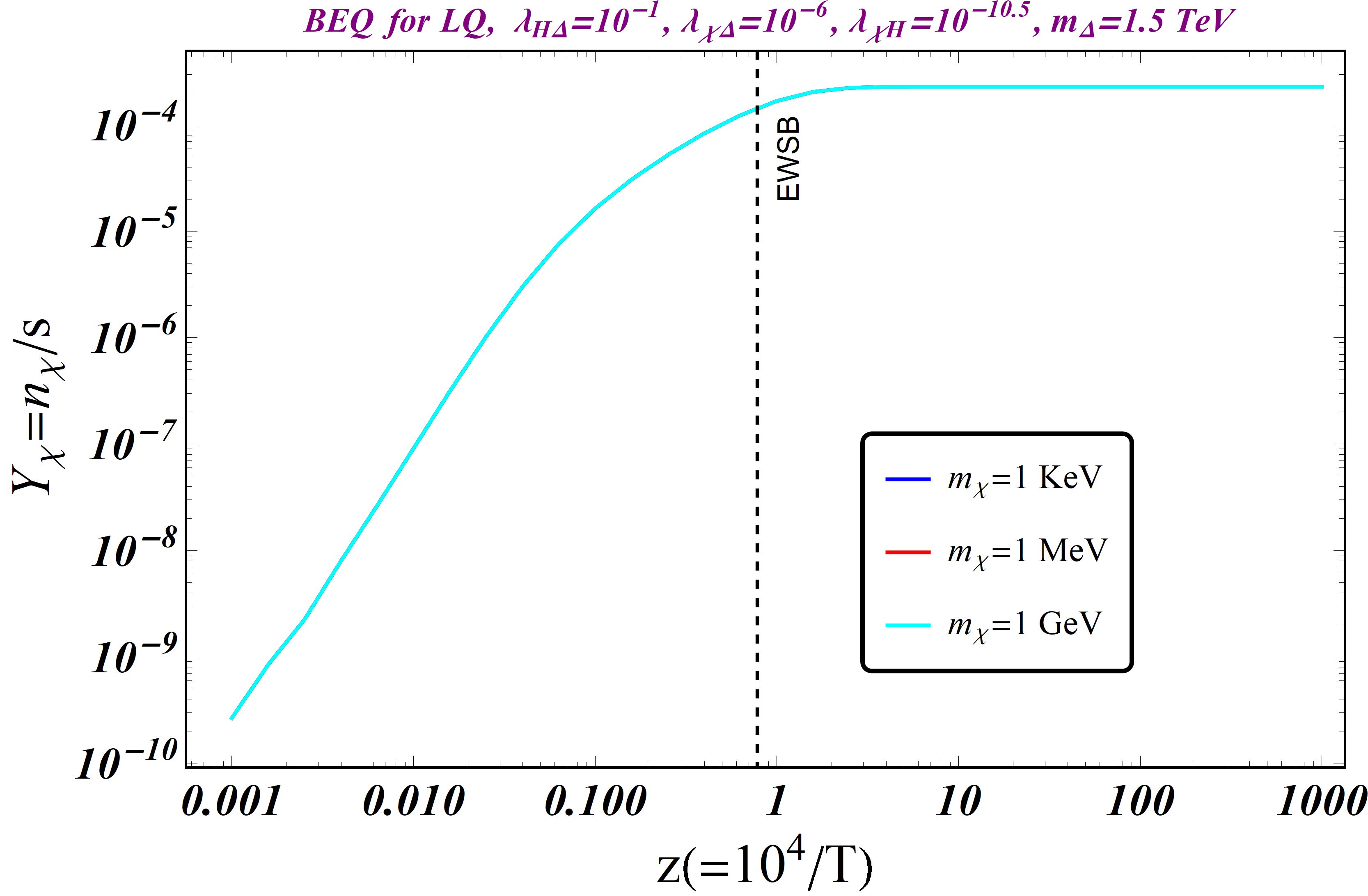}
\caption{DM yield dependence on DM mass}
\label{Fig:LQ DM yield with diff DM masses}
\end{subfigure}
\caption{Dependencies of DM production from LQs on two relevant mass parameters.}
\label{Fig:LQ DM yield 1}
\end{figure}

Finally, the evolution of the dark matter abundance as a function of the temperature for different singlet LQ masses is shown in Fig.~\ref{Fig:LQ DM yield with diff LQ masses}. The general behavior of the DM yield ($Y_{\chi}$) is as expected in the freeze-in scenario,where \(Y_{\chi }\) monotonically increases as the Universe expands and cools, eventually stabilizing and remaining constant once the thermal production rate drops below the Hubble expansion rate. The relevant couplings are chosen as discussed above. Since we find that with heavier LQ mass the DM yield decreases, we choose the LQ mass to be $1.5 \TeV$ for our analysis. 


\textbullet~\textbf{Constraint on $\mchi$:}  

Depending on the scenario chosen (freeze-in or freeze-out) for the DM mass generation, the mass range can vary from $\KeV$ to $\TeV$. Ref. \cite{Bernal:2017kxu} and the references therein provide a good description of different scenarios with such a DM mass range. Absence of a strict upper limit for the DM masses in the freeze-in scenario prompted us to explore the DM yield while varying the DM mass range from a few $\KeV$ to fa ew $\TeV$ for our analysis. The result of that exploration is illustrated in Fig.~\ref{Fig:LQ DM yield with diff DM masses}, and from there we find that the yield of the DM abundance, in the presence of a scalar LQ in the thermal bath, doesn't depend on the dark matter mass.

\begin{figure}[H] 
\captionsetup{justification=raggedright,singlelinecheck=off}
\centering
\includegraphics[width=12 cm]{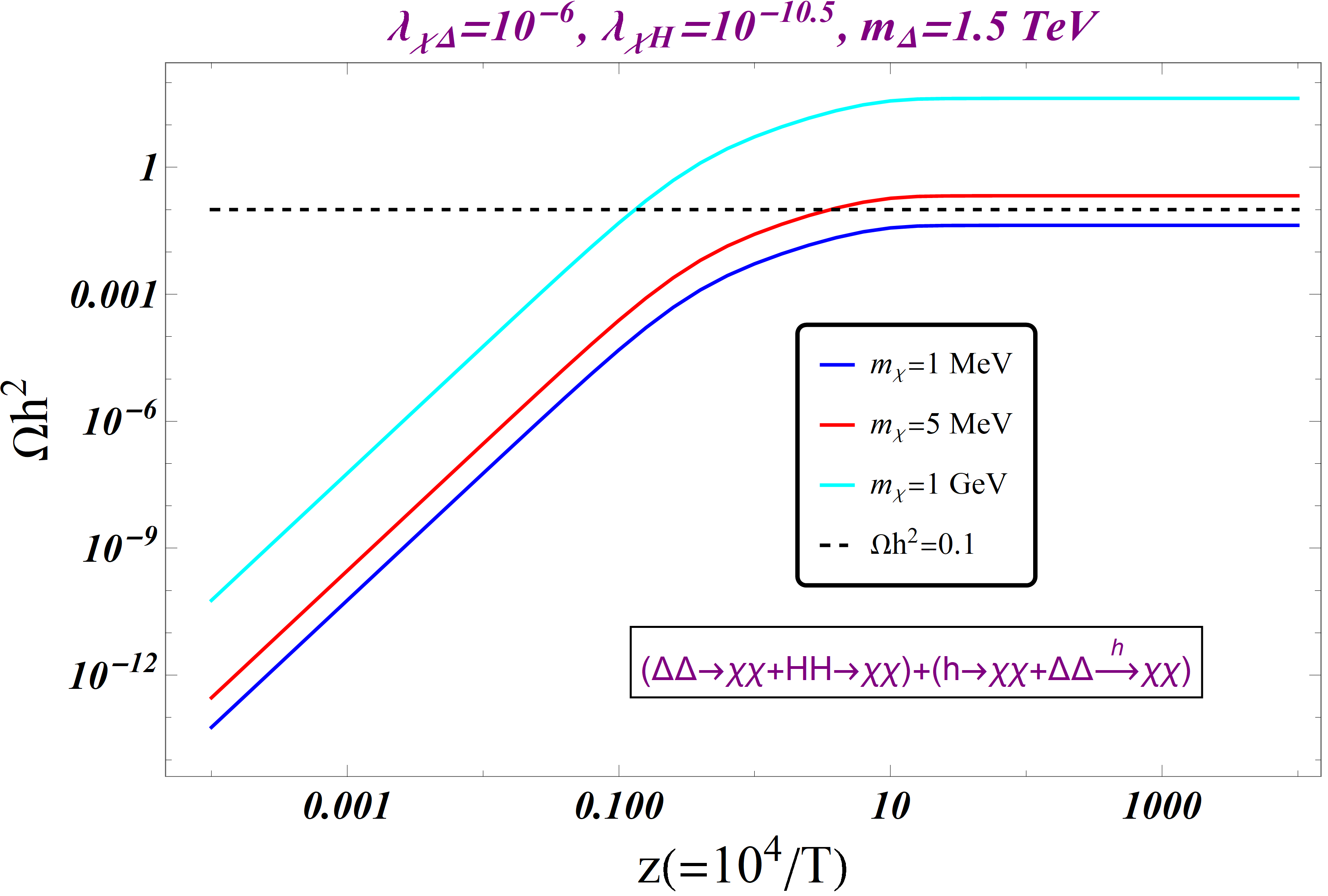}
\caption{The relic density ($\Omega h^2$) generated from the annihilation of leptoquarks and Higgs particles as a function of DM mass, $m_{\chi}$. Other couplings and parameters are set as labeled in the figure.}
\label{Fig:LQ Relic density for diff masses}
\end{figure}

Interestingly, some limit on the DM mass can be obtained from the relic density plot, as shown in Fig.~\ref{Fig:LQ Relic density for diff masses}. This figure shows the total relic density obtained from the Higgs and LQ particles only, both before and after EWSB, for three different DM masses, shown in colors blue ($1 \MeV$), red ($5 \MeV$) and cyan ($1 \GeV$), respectively. The nature of the curves can be understood from Eq.~\ref{eq:Relic density Defn} which tells us that the relic density is directly proportional to the DM mass. As mentioned in the figure, in generating this plot, all dominant processes responsible for generating dark matter, before and after the EWSB, are considered and also mentioned in the legend. From this figure, we also find that to obtain the observed relic density of $\Omega h^2=0.12$ in the present scenario, with other parameters as defined in the figure, the DM should satisfy the condition $m_{\chi}\lesssim 1 \MeV$.

\textbullet~\textbf{Contribution of LQs in the freeze-in process}

It is indeed interesting to analyze the contribution of the LQs in producing dark matter in this present scenario. It has already been discussed that there are three parameters, $m_{\Delta}$, $m_{\chi}$ and $\lamchiD$, which are responsible for the DM yield from the LQs. Therefore, in the following, the dependencies of DM yield on these parameters will be discussed in detail. 

\begin{figure}[H] 
\centering
\includegraphics[width=10 cm]{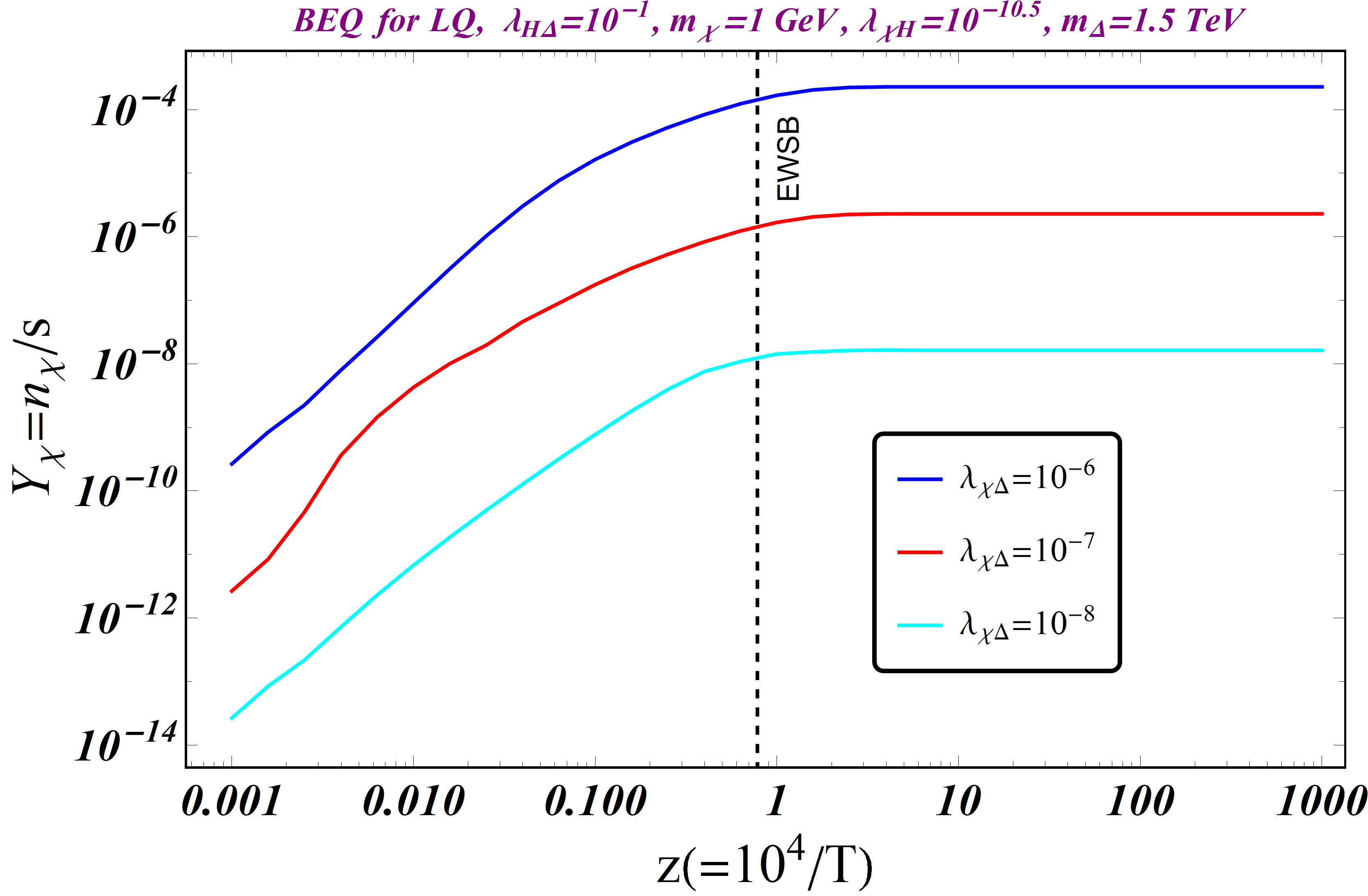}
\caption{DM yield dependence on $\lamchiD$ coupling}
\label{Fig:DM yield with diff LQ-DM coupling}
\end{figure}

Fig.~\ref{Fig:LQ DM yield 1} represents the DM yield dependencies of LQs on two mass parameters. The left panel shows the dependency of the yield on different LQ masses whereas, the right panel shows the same for different DM masses. Fig.~\ref{Fig:LQ DM yield with diff LQ masses} shows the DM abundance for a fixed DM mass of $1 \GeV$ and three different LQ masses of $1.5, 2$ and $2.5 \TeV$, shown in blue, red and cyan color respectively, while keeping the relevant couplings fixed at $\lamchiH=10^{-10.5}$, $ \lamchiD=10^{-6}$ and $\lamHD=10^{-1}$. We find that the yield is reduced for heavier LQ mass, as expected. Similarly, Fig.~\ref{Fig:LQ DM yield with diff DM masses} shows the DM yield for a fixed LQ mass of $1.5 \TeV$ and the same couplings values, but for three different DM masses of $1 \KeV$, $1 \MeV$ and $1 \GeV$. We find almost no change in the DM yield due to these changes in the DM mass. The effect of the DM-LQ coupling, $\lamchiD$, on the DM yield is presented in Fig.~\ref{Fig:DM yield with diff LQ-DM coupling} where we find that the yield gradually decreases with the smaller value of the coupling.

\begin{figure}[t]
\captionsetup{justification=raggedright,singlelinecheck=off}
\centering
\includegraphics[width=11 cm]{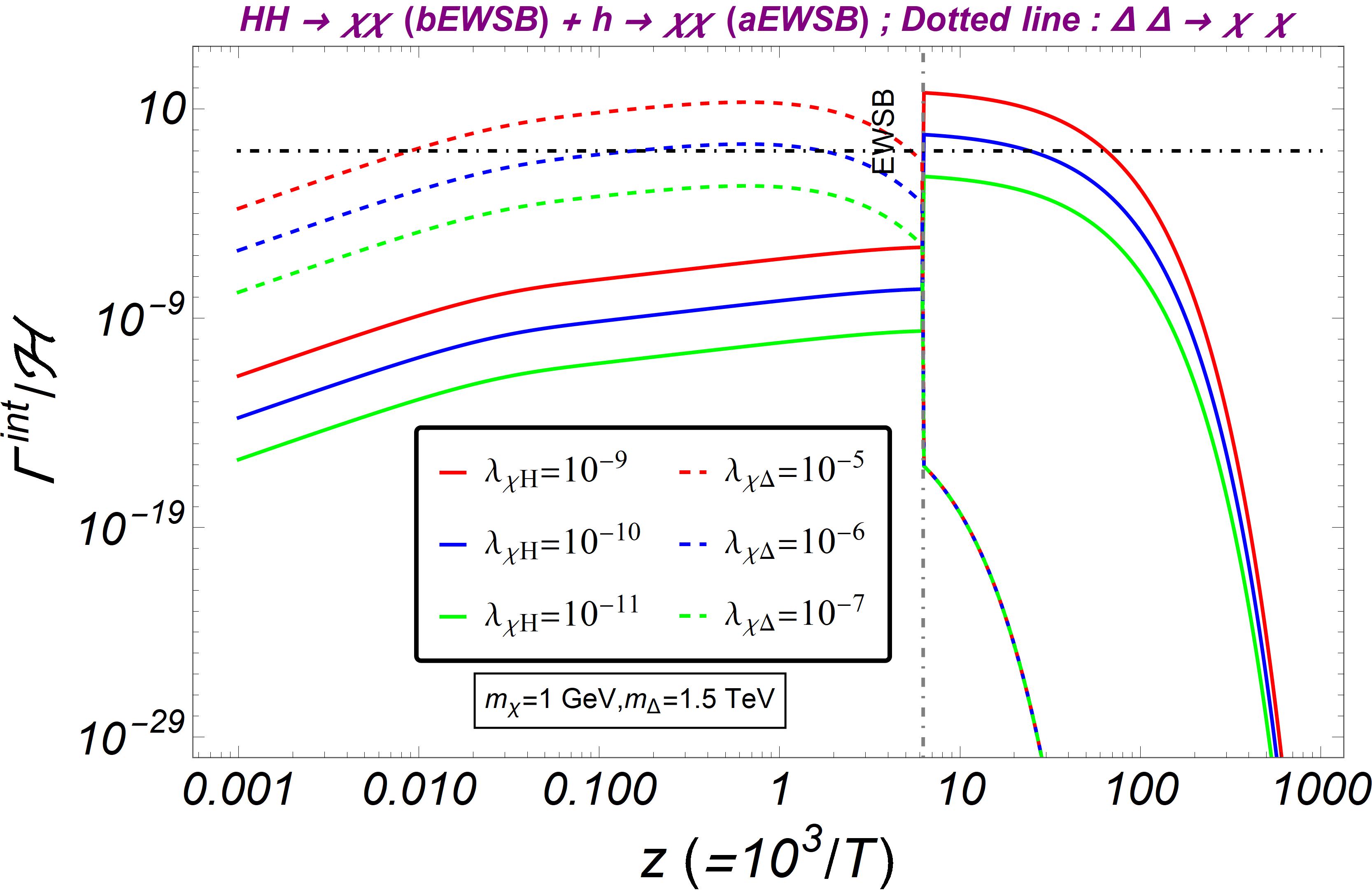}
\caption{A comparative study of the ratio of interaction rate $(\Gamma^{\text{int}})$ and the Hubble parameter $(H)$ for all production channels of DM particles, before and after the EWSB, as a function of inverse temperature (z) for the LQ mass of $1.5 \TeV$, is shown in this plot. The horizontal black dash-dotted line represents the relic density $\Omega h^2=0.1$, whereas the vertical gray dot-dashed line represents the EWSB transition region.}
\label{Fig:Interaction rate comparison}
\end{figure}

\begin{figure}[t] 
\captionsetup{justification=raggedright,singlelinecheck=off}
\centering
\includegraphics[width=11 cm]{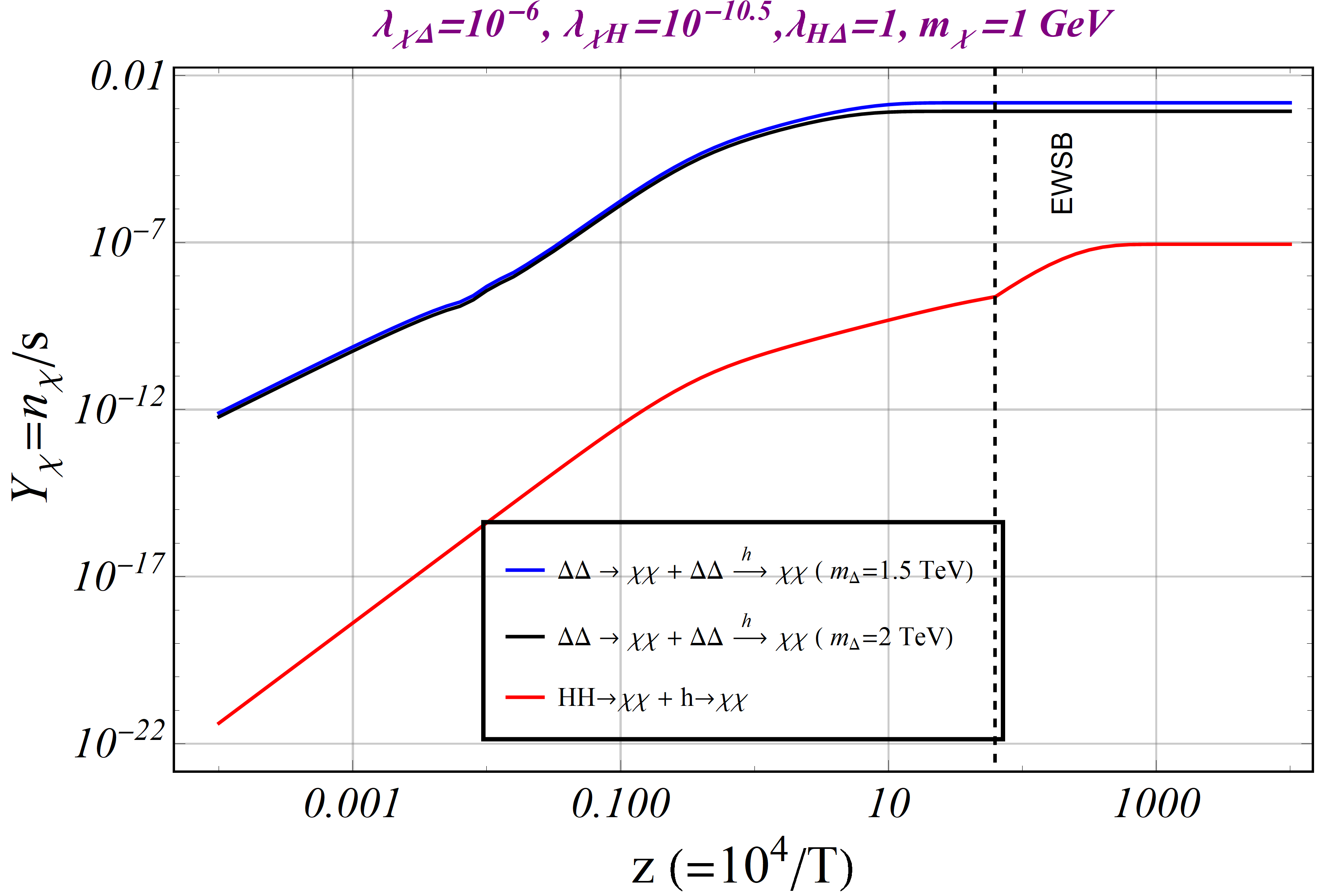}
\caption{A comparative study of the DM yields from the Higgs and the LQ particles. The blue and black lines present the DM yields from the LQs with masses of $1.5 \TeV$ and $2 \TeV$, respectively, whereas the red line presents the yield from the Higgs. All DM producing processes are considered in generating this plot.}
\label{Fig:DM yield comparison}
\end{figure}

\textbullet~\textbf{Comparison between the Higgs and the LQ:}  Being the source of the most dominant DM production mechanisms considered in the present scenario, it is interesting to perform a comparative study between the LQs and the Higgs for relevant processes. These comparisons have been presented in Figures~\ref{Fig:Interaction rate comparison} and \ref{Fig:DM yield comparison}, where the interaction rates among responsible particles to produce the DM and the corresponding yields are shown, respectively.

Solid lines in Fig.~\ref{Fig:Interaction rate comparison} show the total interaction rate (divided by the Hubble parameter) against the inverse temperature $(z)$, for the DM production from the Higgs particle, bEWSB and aEWSB. Different solid colors are for different coupling strengths of $\lamchiH$, as mentioned in the legend. Similarly, the interaction rates for DM production from the LQs with different coupling strengths of $\lamchiD$ are shown by dotted lines in the legend of the figure. The masses of the DM and the LQ were kept fixed at $1~\GeV$ and $1.5~\TeV$, respectively. Total contribution from the Higgs particle comes in two phases, as explained earlier, bEWSB and aEWSB, and thus there is a sharp increment of the interaction rates near the EWSB transition line. Different combinations of couplings are shown by different color codes, as evident from the plot. 

Fig.~\ref{Fig:DM yield comparison} shows the comparative study of the DM production from the Higgs and the LQ particles. As mentioned in the legend, blue and black lines present the DM yields from the LQs with masses of $1.5 \TeV$ and $2 \TeV$, respectively, bEWSB and aEWSB. The slight change in the yield due to the change in the mass of the LQ is already explained above with reference to Fig.~\ref{Fig:LQ DM yield with diff LQ masses}.  The red curve represents the same for the Higgs. Interestingly, we find that the yield from the LQs is almost 4 orders of magnitude higher than the yield obtained from the Higgs.


\textbullet~\textbf{Effect of thermal correction:} An important and interesting aspect of DM production in the freeze-in scenario is the consideration of  thermal masses of different particles involved in the dark matter phenomenology. It has been shown (see Ref.~\cite{Chakrabarty:2022bcn} and discussion therein) that such thermal corrections might open up some dark matter production channels which otherwise are prohibited in the standard freeze-in scenario. Therefore, to be consistent with this analysis, we did check the thermal effect on the particles involved in this analysis for the DM production. 

As explained before, it is a known fact that due to the suppression by propagators and tiny couplings, the annihilation processes in the freeze-in mechanism are significantly less contributing (compared to the decay processes) towards the DM production. Since only the Higgs particle is responsible for the DM production through decay, we shall very briefly discuss the thermal effect on it and neglect such analysis for the LQ particle.

\begin{figure}
\captionsetup{justification=raggedright,singlelinecheck=off}
\centering
\includegraphics[width=11 cm]{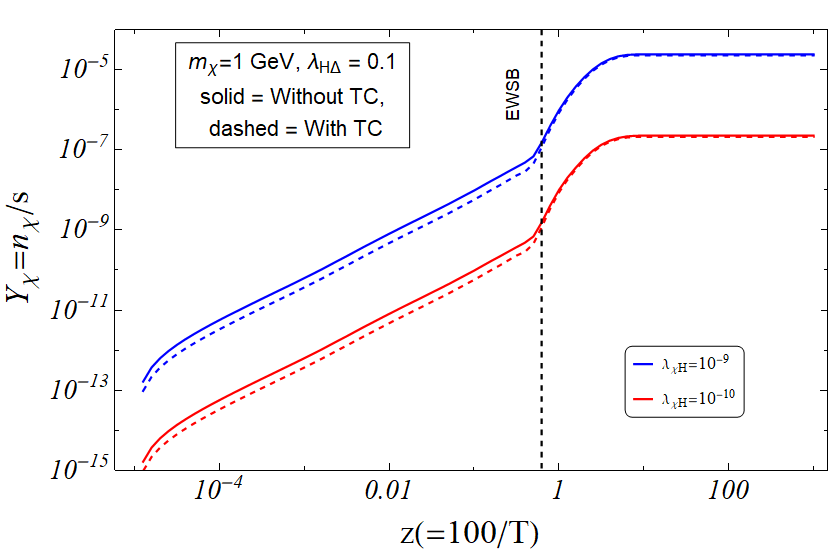}
\caption{A comparative study of the total DM yield from the Higgs particle only (including both, scattering and decay), with (dotted lines) and without thermal correction (solid lines), is presented in this figure. Different colors represent different coupling strengths $(\lamchiH)$ between the DM and the Higgs particle.}
\label{Fig:DM yield comparison with TC}
\end{figure}

Fig.~\ref{Fig:DM yield comparison with TC} represents a comparative study of the relic density generation from the Higgs particle, with (dotted lines) and without thermal correction (solid lines). The solid blue curve shows the DM yield for the coupling $\lamchiH=10^{-9}$, at temperature $T=0$, whereas the dotted blue curve shows the same for $T\neq 0$. The formula used for this thermal effect calculation is provided in the Appendix (\ref{eq: Higgs thermal corrcn 2}).  From that equation we find that the thermal correction to the Higgs mass, $\delta m_h^2(T)$, is directly proportional to the self-coupling of the Higgs $\lambda_H$ and the Higgs-DM coupling, $\lamHD$. Among these, $\lambda_H$ is not relevant for the DM production and  $\lamHD$ is responsible for the Higgs-mediated LQ scattering process only. Keeping this coupling fixed at 0.1 and with a DM mass of $1 \GeV$, the plot was obtained for two different values of $\lamchiH$, $10^{-9}$ (blue) and $10^{-10}$ (green). Although it is not presented, similar results are obtained for the values of 1 or even 5. From the nature of this plot, it can be clearly concluded that the thermal correction is not playing any significant role in the present analysis.


\section{Conclusion}
\label{Sec:conclusion}

In this article, a simple analysis is performed in the freeze-in scenario, in the presence of a scalar LQ and a scalar DM which are neutral and odd, respectively, under the $Z_2$ symmetry. The main purpose of this analysis was to estimate an
order-of-magnitude DM yield for the scenario, rather than a very precise calculation. The parameters responsible for this setup were identified and their ranges were analyzed. The main conclusions that can be summarized from this analysis are the following. The presence of a heavy particle like Leptoquark in the thermal bath has some significant effect on the DM yield in the freeze-in scenario, as can be seen in Fig.~\ref{Fig:DM yield comparison}. Without any known bound on the Higgs-LQ coupling $(\lamHD)$, it can be taken of the order of unity. From Fig.~\ref{Fig:LQ-DM coupling range}, we find that the LQ-DM coupling in this setup should be $\lamchiD\lesssim 10^{-6}$. Similarly, from Fig.~\ref{Fig:Higgs coupling bEWSB aEWSB}, it was found that the Higgs-DM coupling, $\lamHD$, in this setup should be taken as $\lamchiH\lesssim 10^{-10.5}$. This small coupling is a well-known fact and is one of the salient features of the freeze-in scenario. To obtain the observed relic density of $\Omega h^2=0.12$, a DM should satisfy the condition $m_{\chi}\lesssim 1 \MeV$, as evident from Fig.~\ref{Fig:LQ Relic density for diff masses}. Leptoquarks' contribution to the DM yield is mainly affected by the $\lamchiD$ coupling while being almost independent of the two mass parameters, as can be seen from Figs.~\ref{Fig:LQ DM yield 1}-\ref{Fig:DM yield with diff LQ-DM coupling}. Since the relevant couplings are extremely small, they easily evade the current and future direct detection experiments like LUX-ZEPLIN (LZ) or similar ones.

\section*{Acknowledgment}
I sincerely thank Purusottam Ghosh for valuable discussions and providing calculation methods to perform the analysis.


\appendix

\section{Other annihilation cross sections for scalar dark matter}
\label{appendix:other xscns}

\begin{enumerate}
    \item $f{\bar f} \rightarrow \chi\chi $ with $f$ being the SM fermions and $\chi$ being the dark matter,

\be \label{eq:fftoDM}
\sigma_{f{\bar f} \rightarrow \chi\chi} = \frac{\lamchiH^2 m_f^2 N_c}{8 \pi  s}\frac{\sqrt{\left(s-4 m_f^2\right) \left(s-4 m_{\chi} ^2\right)}}{ \left(s-m_h^2\right)^2 + m_h^2\Gamma_h^2}.
\ee

\item $V_1V_2\rightarrow \chi\chi$ with $V$ being electroweak gauge bosons, $W$ and $Z$

\be \label{eq:VVtoDM}
\sigma_{V_1 V_2 \rightarrow \chi\chi} = \frac{\lamchiH^2 }{1052 \pi}\sqrt{\frac{s-4 m_{\chi}^2}{s-4 m_V^2}}\frac{s}{ \left(s-m_h^2\right)^2}\left(2+\frac{12 m_V^4}{s^2}-\frac{4 m_V^2}{s}\right) 
\ee

\begin{figure}[H] 
\begin{center}
\resizebox{3cm}{3cm}{
\begin{tikzpicture}
\begin{feynman}
\vertex (a) at (0, 0);
\vertex (b) at (-2, 1) {\(f\)};
\vertex (c) at (-2, -1) {\(\bar{f}\)};
\vertex (d) at (2, 0);
\vertex (e) at (4, 1) {\(\chi\)};
\vertex (f) at (4, -1) {\(\chi\)};
\diagram* {
(b) -- [fermion] (a) -- [fermion] (c),
(a) -- [scalar, edge label'=\(h\)] (d),
(e) -- [scalar] (d) -- [scalar] (f),
};
\end{feynman}
\end{tikzpicture}}\quad\quad\quad\quad
\resizebox{3cm}{3cm}{
\begin{tikzpicture}
\begin{feynman}
\vertex (a) at (0, 0);
\vertex (b) at (-2, 1) {\(V\)};
\vertex (c) at (-2, -1) {\(\bar{V}\)};
\vertex (d) at (2, 0);
\vertex (e) at (4, 1) {\(\chi\)};
\vertex (f) at (4, -1) {\(\chi\)};
\diagram* {
(b) -- [boson] (a) -- [boson] (c),
(a) -- [scalar, edge label'=\(h\)] (d),
(e) -- [scalar] (d) -- [scalar] (f),
};
\end{feynman}
\end{tikzpicture}}
\end{center}
\caption{Higgs mediated tree-level Feynman diagrams for annihilations of fermions and gauge bosons to scalar dark matter. $V$ stands for $W$ and $Z$ gauge boson.}
\label{Fig:DM production from annihiln}
\end{figure}
\end{enumerate}


\section{Thermal corrections} \label{sec:Thermal Correction}

As the Higgs is the only particle getting VEV in our setup, we shall consider the thermal correction to the Higgs mass only. At temperature $T$ this is given by \cite{DeRomeri:2020wng}
\be \label{eq: Higgs thermal corrcn 1}
m_h^2 \rightarrow m_h^2 + \delta m_h^2(T)
\ee
where 
\bea \label{eq: Higgs thermal corrcn 2}
 \delta m_h^2(T) &=& \Big(\frac{1}{16}g^{\prime 2} + \frac{3}{16}g^2 + \frac{1}{4}y_t^2
 + \frac{1}{2}\lambda_H + \frac{1}{12}\lamHD\Big)T^2\,.
\eea
In Eqn. \ref{eq: Higgs thermal corrcn 2}, $g^{\prime},~g$ and $g_s$ are the $U(1)_Y$, $SU(2)_L$ and $SU(3)_C$ gauge couplings, $y_t$ is the Yukawa coupling of the top quark and $\lambda_H$ is the Higgs self-coupling taken as 0.12.

\bibliographystyle{JHEPCust.bst}
\bibliography{SLQSDM.bib}

\end{document}